\documentclass[10pt]{article}

\usepackage{authblk}

\usepackage{ifpdf}
\usepackage{graphicx,amssymb,amsmath}
\usepackage{natbib}
\usepackage{latexsym,theorem}
\usepackage{amssymb,amsmath}
\usepackage{euscript}
\usepackage[english]{babel}

\usepackage[utf8]{inputenc}
\usepackage[english]{babel}

\usepackage{natbib}

\theoremstyle{plain}
\newtheorem{theorem}{Theorem}

{\theorembodyfont{\upshape}
\newtheorem{definition}[theorem]{Definition}

\newtheorem{lemma}[theorem]{Lemma}

\usepackage{xcolor}   

\newcommand{\RR}{{\mathbb R}} %
\newcommand{\EE}{{\mathbb E}} %
\newcommand{\PP}{{\mathbb P}} %

\newcommand{\QQ}{{\mathbb Q}} %

\def\text#1{\quad\mbox{#1}\quad} %wide text in mathsok
\def\defegal{\triangleq}

\def\F{\mathcal{F}}

\newenvironment{preuve}{\small{\bf
    Proof:}}{\hfill$\Box$\normalsize\\\bigskip}

\newenvironment{keywords}{\small{\bf
    Key Words:}}{\hfill\normalsize\\ \bigskip}

\title{Regime Switching Entropic Risk Measures on Crude Oil Pricing}  
\author{Babacar Seck$^{1}$ and Robert J.  Elliott$^{2}$ \\
{}$^{1}$bseck@uob.edu.bh,\;{}$^{2}$relliott@ucalgary.ca
}

\begin{document}
\maketitle
\date{}
%\setcounter{tocdepth}{2} %LC-35

%\tableofcontents

\vspace{0.5cm}
\begin{abstract} 
% 1. Research Problem; 2. Methods; 3. Results; Conclusion
This paper introduces a new type of  risk measures, namely regime switching entropic risk measures, and study their applicability through simulations.
 The state of the economy is incorporated into the entropic risk formulation by using a Markov chain. Closed formulae  of the risk measure are obtained for futures on crude oil derivatives. The applicability of these new types of risk measures is based on the study of the risk aversion parameter and the 
 convenience yield. %The comparison analysis is based on the risk aversion parameter of the entropic risk measure, the convenience yield of the derivative on crude oil and their behavior over time. 
 The numerical results show a term structure and a mean-reverting behavior of the convenience yield.
\end{abstract}


%\section{introduction and Motivation}
\section{Introduction}
\label{Introduction}
Often, the state of the economy is described as good (growing economy) or bad (recession). Each of these two states has a very specific meaning and the description is based on some key indicators such as: inflation, unemployment rate, housing markets, consumer spending, confidence level of householders and so one. 
Stock markets and commodity prices  depend on these performance indicators, which ultimately determine the gross supply and demand of any good or service.  However, technical indicators developed by direct actors 
can be used to describe and predict market dynamics as illustrated in~\cite{Chris2014} and in~\cite{Rashid2008}.
Some other key factors that may influence supply and demand are the risk environment and economic downturns.
The  risk environment seems to have  direct impacts on the energy market. Thus, risk  aversion, loss aversion and preferences can be considered as consequences of the macroeconomic, microeconomic  indicators mentioned above and determine the pricing in any given market, in particular the crude oil market.  Preferences and aversions can be time dependent, stable or time consistent. These terminologies have specific meanings and describe the decision maker's behaviour over time. However,  theoretical and experimental measurements for these indicators lead to different results and diverge sometimes,  see~\cite{Anderson2009},~\cite{Hannah2018},~\cite{Dohmen2016},~\cite{Gollier2001} for further discussions. 
 The list is not exhaustive; some references therein might be worth  considering.  Now, the challenge is to choose a model 
to capture both the market dynamics and the decision maker's behaviour. To this end, we use three levels of modelling to quantify the risk associated  with future contracts and swaps on the crude oil markets: the pricing processes itself, the behaviour of the decision maker (utility functions and aversion) and the market regulations (risk measures and capital requirement). In this study, we consider the first two levels of modelling.   \\[3mm]

Following the seminal work of~\cite{Hamilton1989}, a finite state Markov chain can be introduced into  mathematical finance to model the states of an economy which  affect the price of any contingent claim. We derive some closed form expressions related  to entropic risk utility functions. 
Entropic risk measures are a class of utility based  risk measures  that extend the notion of coherent risk measures   first introduced in~\cite{Delbaen1997,Delbaen1999}.  They are based on the concept of relative entropy, which is a measure of divergence between two probability distributions\footnote{There are different interpretations of divergence of probability measures, most of them  follow the seminal work of Kullback and Leibler divergence, see~\cite{Kullback1951}  }. Two classes of entropic risk measures have been compared in~\cite{Mario2018}, namely coherent or convex. These properties of risk measures are discussed largely in the literature and express some nice properties in terms of decision making strategies or solving complex portfolio optimization problems, see~\cite{Delbaen2006bis},~\cite{Detlefsen2005},~\cite{Teboulle:util2},~\cite{Shapiro2004},~\cite{Follmer2006},~\cite{Seck2011}. 
Entropic risk measures have an exponential function representation. Moreover, coherent entropic risk measures can be expressed  in term of convex entropic risk measures. Thus, we shall focus on the exponential formulation of entropic risk measures to derive closed form formulae that quantify the risk associated with a give contingent claim.
%We shall also use a utility function to express loss aversion for their simplicity of formulation and intuitive interpretation of the decision maker's behaviour. 
See~\cite{Goovaerts2004} where the entropic risk measure is introduced by using an exponential utility function.
 The entropic risk measures we formulate are time consistent as they are expressed using the conditional expectation operator. They depend  on the state of an economy and could be thought  as the price of the  contingent claim in an incomplete market.
This paper is organized as follow. In Section~\ref{market-preferences}, we set up the framework used to model the market dynamics that influence the price of the risky asset and the preferences of a rational  decision  maker. Section~\ref{Comparative-study} derives closed form expressions for entropic risk measures in the
Section~\ref{market-preferences} setting. Section~\ref{Sec:Illust} provides a comparison analysis of the regime switching entropic risk measures computed for different crude oil derivatives obtained in Section~\ref{Comparative-study}.
% The comparison analysis is based on the risk aversion parameter of the entropic risk measure, the convenience yield of the derivative on crude oil and their behavior over time.

\section{Market model and decision makers preferences}
\label{market-preferences}

\subsection{Market, price dynamics and preferences}

 Suppose $(\Omega,\F,\PP)$ is a complete probability space, where $\PP$ is a real-world probability.
  Let  $Z=\{ Z_{t}, t \geq 0 \} $ be a continuous-time, finite state homogeneous Markov chain with state space $\mathcal{Z}\defegal
\{z_{1},\ldots,z_{N}  \}  $.
  The state space  $\mathcal{Z}$ is interpreted as different states of an economy.
  Following~\citep{Elliott1994}, we shall identify the state space $\mathcal{Z}$ with the set of unit vectors 
  $\mathcal{E}\defegal \{e_{1},\ldots,e_{N}  \}$ where  $e_{i}=(0,\ldots,0,1,0,\ldots,0)'\in\RR^{N} $ (the $i$th component of the vector 
  is equal to $1$) and
  $'$ represents the transpose of a matrix or
a vector. 
$Z$ then has the following semi-martingale decomposition:
 \begin{equation}
 \label{markovchaine}
 Z_{t}=Z_{0} + \int_{0}^{t} AZ_{u}\mathrm{d}u+M_{t}\in\RR^{N}\,.
 \end{equation}
% see~\citep{Elliott1994}.
% where $A=[a_{ji}]_{i,j=1,\ldots,N}$ and $a_{ji}=\PP\big( Z_{t}=e_{j}|Z_{t-1}=e_{i}\big)$.
%$A$ is the transition matrix of the Markov chain and $(M_{t})_{t\in(0,T]}$ is a martingale increment
 %with respect to the filtration $\F^{Z}$ generated by $Z$.
Here  $A=[a_{ij}]_{i,j=1,\ldots,N}$ is the rate matrix of the Markov chain and $(M_{t})_{t\in(0,T]}$ is a martingale
 with respect to the filtration $\F^{Z}$ generated by $Z$.

Suppose that the spot price of a risky asset $X$ in the market is given by 
an Ornstein-Uhlenbeck process  satisfying the stochastic differential equation:
\begin{equation}
\mathrm{d}X_{t}=\alpha(\mu-X_{t})\mathrm{d}t +\sigma\mathrm{d}B_{t}.
\end{equation}
Here $X_{0}=x_{0}$ is the initial value or condition of the process. The parameters $\alpha,\mu,\sigma\in\RR$ are deterministic, where
$\alpha$ is the reversion rate, $\mu$ is the mean reversion level or the long-run mean value of the risky asset (it can be interpreted as an equilibrium value) and $\sigma$ the long-term standard deviation or volatility. We denote by
 $B_{t}$  a standard Brownian motion. The above one factor mean-reverting model can be extended in such a way that
 the mean-reversion level becomes a function of time $\mu(t)$. Nevertheless, it will not change the closed form expressions for the path of the stochastic process nor the level of difficulty of the numerical simulations and the parameter estimates, see~\cite{Marin2016}.
 
 By applying
 Ito's Lemma to the function $(x_{t},t)\mapsto x_{t}e^{\alpha t}$, we obtain:
\begin{equation}
\label{EDS1}
X_{t}=x_{0}e^{-\alpha t} + \mu(1-e^{-\alpha t})+\sigma\int_{0}^{t}e^{\alpha (u-t)}\mathrm{d}B_{u}.
\end{equation}
Knowing $X_{s}$, $0\leq s <  t $, $X_{t}$ is a  Normal distribution $\mathcal{N}(\mu_{s,t},\sigma_{s,t})$, where
\begin{equation}
\mu_{s,t}=  x_{0}e^{-\alpha (t-s) } + \mu(1-e^{-\alpha (t-s)})  \quad \mbox{and}  
\quad \sigma_{s,t}=  \frac{\sigma^{2}}{2\alpha}(1-e^{-2\alpha (t-s) }).
\end{equation}
%and for $0\leq s < t \leq T $
%\begin{equation}
%\EE[X_{t}\mid X_{s}]\defegal\mu_{s,t}=  x_{0}e^{-\alpha (t-s) } + \mu(1-e^{-\alpha (t-s)})  \quad \mbox{and}  \quad 
%\sigma_{t}=  \frac{\sigma^{2}}{2\alpha}(1-e^{-2\alpha t }),
%\end{equation}
\cite{Schwartz1997} used  this one-factor mean reverting stochastic process to model the  spot price
 of a commodity. Here, we shall focus on crude oil spot prices. By far, it is the most traded commodity in the physical and in the future markets. 
 Different factors can explain the spot price of crude oil: 
 \begin{itemize}
\item[\textendash]  fundamentals of supply and demand,
\item[\textendash] geopolitical instability in the Middle East, 
\item[\textendash]  competition of existing energy sources, 
\item[\textendash]  new technological developments less dependent on oil, 
\item[\textendash] restrictions on fossil fuel consumption due to environmental concerns, 
\item[\textendash] the role of OPEC (Organization of the Petroleum Exporting Countries) which makes political decisions in the interests of its members, 
\item[\textendash]  the geographical location of the market,
\end{itemize}
and so on, see~\cite{Baumeister2016}. Thus, it would be unrealistic to choose a specific stochastic model that will model all of these drivers. Based on these observations and the remarks developed in Section~\ref{Introduction}, we shall combine market dynamics and the state of the economy to express contingent claims on the risky asset $X_t$. Closed form expressions for the market risk associated to the contingent claims will de derived.
 
 \subsection{Examples of spot prices, contingent claims and interpretation}
\label{sec:EnergyDerivatives}

\subsubsection{Linear functions of spot prices} 
Let $\mathcal{S}$ be  the linear space of  measurable functions on $\Omega$, including constant functions. Assume that we want to forecast the spot price of crude oil using the model described in~\ref{EDS1}. We know that this framework does not capture all the dynamics that influence the crude oil spot prices. 
Thus, we can attempt to model all the macro economic factors using one single dynamic, the state of the economy $Z_t$. 

We can formulate the value of a contingent claim $S_t$ that belongs to   $\mathcal{S}$ in the form:
%Suppose that the contingent claim $S_{t}$ belongs to $\mathcal{S}$ and  is given by 
\begin{equation}
\label{LinearDerivative}
S_{t}=f(X_{t},Z_{t})\defegal X_{t} < \delta,Z_{t} >, \quad  t=1,\ldots,T,\quad\mbox{where}\quad \delta\defegal(\delta_{1},\ldots,\delta_{N})\in\RR^{N}.
\end{equation}
The above expression of the contingent claim indicates that its value is a linear function of the spot price. The constant real numbers $\delta_i$ stand for the coefficient of linearity between the spot price and its contingent claim, and depend on the state of the economy. In the exchange market, it is actually a proportion of the spot price. However, the linear dependence  between the two variables is not necessarily justified. Next, we will present the most common contracts traded by producers and investors.
\nocite{Andre2013}

%Data website: https://www.eia.gov/finance/markets/crudeoil/financial_markets.php
% Future Contracts Explained MERCATUS https://www.mercatusenergy.com/blog/bid/86597/the-fundamentals-of-oil-gas-hedging-futures

%\\[2mm]

\subsubsection{Future contracts} 
%Baumeister, C. & Kilian, L. (2016). Forty years of oil price fluctuations: Why the price of oil may still surprise us. The Journal of Economic Perspectives, 30, 139?160.
In the context of this regime switching model, we propose a generalization of future contracts on the underlying 
 $X_t$. From  the formulation in Veld-Merkoulova and Roon~(\cite{Veld2003}), we derive the general form:
%Following Veld-Merkoulova and Roon~(\cite{Veld2003}), the price of the forward contracts on $X_{t}$ can be written in the general form:
\begin{equation}
\label{comprices}
\forall\, F_{t}\in\mathcal{S},\quad  F_{t}=f(X_{t},Z_{t})\defegal e^{Y^{Z}_t(c_t,r_t)(t-T)}  X_{t}, \quad  t=1,\ldots,T.
\end{equation} 
where $Y^{Z}_t$ is the convenience yield\footnote{Physical ownership of commodities carries an associated flow of services, known as cost of carry. In the mean time, the owner may benefit from having a direct access to the commodity. The net difference, benefit of ownership minus cost of carry, defines the convenience yield.} over the life of the contract. The convenience yield provides a benefit from holding the underline commodity (namely barrels of crude oil) over a period of time. We assume that it depends on the state of the economy $Z$ and is  a function of  two variables. These variables are:
% $d\defegal(U,I)$ and $v\defegal(r,u,y)$ with:
\begin{itemize}
\item  $r_t$ the continuously compounded risk free rate of return and
\item  $c_t$  the cost of storage. % (proportional to the price of the commodity)
%\item $Y^{Z}_t$ is the convenience yield over the life of the contract, as the convenience yield provides a benefit from holding
%the underline commodity, (crude oil, gold, etc); we assume that it depends on the state of the economy 
\end{itemize}
Attempting to express the convenience yield by using a regime switching approach is relatively new. Meanwhile, the fact that the convenience yield is affected by changes in economic cycles (during economic recovery, inventories are low, and convenience yield is high) is well known. In~\cite{ConvenienceYield3}, the author models the convenience yield as a mean-reverting OU stochastic process and then discretizes this process based on  transition probabilities. This models the mean-reverting behavior of the convenience yield back to the Schwartz model (\cite{Schwartz1997}).
Hence the 
regime switching approach does not have an economic interpretation. Different interpretations of the convenience yield can be found in~\cite{ConvenienceYield2}. 
Note that, the convenience yields are not observable variables. One of the main difficulties is to estimate the value of the convenience yield curve with respect to the term structure of the future contracts.  Thus, a filtering approach might be suitable  as suggested in~\cite{ConvenienceYield1}.  However, in theory the cost of storage can be determined by taking the difference between quoted prices
for two different dates of delivery. We will return to these numerical and pricing aspects of the convenience yield curve
 in Section~\ref{Sec:Illust}.  \\[1mm]
 
\subsubsection{Commodity Swap}
%https://www.e-education.psu.edu/ebf301/node/548
We recall that a commodity swap does not involve physical transactions, instead only cash settlements are observed.
Assume that $\displaystyle \widetilde{F}_t$ is the forward price on the commodity $X_t$ at the beginning of the time period $t$. A simple arbitrage reasoning  allows  to determine  the value of the commodity swap on one unit of commodity as:
\begin{equation}
\label{eq:comswap}
W_T=\sum_{t=1}^{T} e^{-r_t}\Big( \widetilde{F}_t - X_t\Big). 
\end{equation}
Assume that the interest rates are deterministic. In that particular case, the values of the future and froward contracts coincide. That is 
\begin{equation}
\widetilde{F}_t  =  F_{t}\defegal e^{Y^{Z}_t(c_t,r_t)(t-T)}  X_{t}, \quad  t=1,\ldots,T. 
\end{equation}

If we substitute the above equation in~\eqref{eq:comswap}, we derive explicitly the value of the commodity swap as follows:
\begin{equation}
\label{eq:comsetl}
W_T= f(X_t, Z_t)=\sum_{t=1}^{T} e^{-r_t}X_t\Big( e^{Y^{Z}_t(c_t,r_t)(t-T)}   - 1 \Big).
\end{equation}
We can interpret the above formula. To do so, write $w_t$ the discounted cash settlement of the swap for the time period $[t-1,\, t)$. That is 
$w_t= e^{-r_t} X_t\Big( e^{Y^{Z}_t(c_t,r_t)(t-T)} - 1\Big)$ and the cash flow at time $t$ is $ e^{r_t}w_t=  X_t\Big( e^{Y^{Z}_t(c_t,r_t)(t-T)} -1\Big)$.
\begin{itemize}
\item If $X_t>0$, the cash settlement and  $e^{Y^{Z}_t(c_t,r_t)(t-T)} - 1$ have the same sign. Then,  from the producer perspective:
\begin{itemize} 
\item a credit on the time period $[t-1,\, t) $ means that holding one unit of the commodity will generate a profit of $e^{Y^{Z}_t(c_t,r_t)(t-T)} - 1 $. Then it is of the interest for the producer to maintain or increase his inventory level rather than holding cash.
\item a debit on the time period $[t-1,\, t)$ means  that, the producer should sell the underlying commodity and decrease his inventory. One can make this deduction by using the opposite of the above reasoning.
\end{itemize}
\item If $X_t< 0$, we are dealing with an extreme and rare situation as market prices of goods are in general positive. In that case, the cash  settlement $w_t$ and  $e^{Y^{Z}_t(c_t,r_t)(t-T)} - 1 $ have opposite signs. Then, the two following situations can happen:
\begin{itemize}
\item  a credit on the time period $[t-1,\, t) $ means that $e^{Y^{Z}_t(c_t,r_t)(t-T)} - 1 < 0$. That is the producer should reduce his inventory and sell the underlying asset even if the spot price is quoted negative. This is what happen at  the beginning of the Covid-19 pandemic in North America on April 2020, when the May WTI crude oil closed at $-\$ 37.63$ US dollars a barrel. Physical transactions of the underlying were then involved.\footnote{This is exactly the opposite so-called short squeeze phenomenon. In a short squeeze, traders that are on short position fear they will be unable to find the underlying physical commodity and are forced to cover their positions, driving prices up sharply.} The demand dropped and producers were over stocked.
\item  a debit on the time period $[t-1,\, t) $ means that $e^{Y^{Z}_t(c_t,r_t)(t-T)} - 1 > 0$. Then the producer should increase his inventory and refrain from selling at the spot price. Hence a short squeeze phenomenon in the future market can help to bring the spot price to the positive side.
\end{itemize}
\end{itemize}
Stochastic models that caste the dynamic of the convenience yield has been the object of several works; among them the Gibson-Swartz model. Next, we shall formulate this model in the context of a Markov modulated setting.  \\
\emph{An example of a stochastic convenience yield:}
Following the Gibson-Swartz model, a simple way to express the dynamics of the convenience as  a function of the state of the economy is:
\begin{equation}
\label{eq:YZ}
Y^{Z}_t(c_t,r_t) = Y_t(c_t,r_t) <\delta,Z_t >.
\end{equation}
The dynamic of $Y_t(c_t,r_t)$ is given by the the OU process
\begin{eqnarray}
\mathrm{d}X_{t} & = & (r_t - Y_{t})X_t\mathrm{d}t +\sigma_X X_t\mathrm{d}B^1_{t},\\
\mathrm{d}Y_{t} & = & \kappa (y - Y_{t})\mathrm{d}t +\sigma_Y \mathrm{d}B^2_{t},
\end{eqnarray}

where $B^1$ and $B^2$ are $1$-dimensional Wiener processes such that $\displaystyle <B^1, B^2 >_t=\rho \mathrm{d}t$, under the risk neutral probability $\QQ$, see~\cite{GS90}.
 In this model, we assume that the cost of carry $c_t=0$ and the interest rate $r_t$ is deterministic. Numerical simulations of  OU processes is widely used in the literature
and reasonable values for $\rho$ have been determined. We should mention that, this model is developed under a risk neutral probability.
The estimation of the parameters and the risk associated to any contingent claims will be based on the observations, i.e. the  historical probability. Thus, the historical dynamics of the convenience yield can be expressed as
\begin{equation}
\label{eq:DymanicY}
\mathrm{d}Y_{t}  =  \kappa (\widehat{y} - Y_{t} - \lambda_Y)\mathrm{d}t +\sigma_Y \mathrm{d}\widetilde{B}^2_{t},
\end{equation}
for some bivariate Brownian motion $ \Big( \widetilde{B}^1, \widetilde{B}^2 \Big)$ under the historical probability $\PP$, 
 with  the  same correlation as 
$ \Big( B^1, B^2 \Big)$ under $\QQ$.  The equation is the same as before, as long as we
replace $y$ by  $ \widehat{y} = \displaystyle  y - \frac{\lambda_Y}{\kappa} $.
 Write $\widetilde{Y}_t\defegal  \kappa (\widehat{y} - Y_{t} - \lambda_Y)$, the expression for the dynamics of the convenience yield \eqref{eq:DymanicY}
can be simplified as
\begin{equation}
\mathrm{d}Y_{t}  = \widetilde{Y}_t  \mathrm{d}t +\sigma_Y \mathrm{d}\widetilde{B}^2_{t}.
 \end{equation}
Then, we can determine the dynamic of the  Markov modulated convenience yield formulated in \eqref{eq:YZ}.
By definition $Y^{Z}_t=Y_t <\delta,Z_t >= <\delta, Y_t \cdot Z_t >$. Then $\mathrm{d}Y^{Z}_{t} = <\delta, \mathrm{d}\big(Y_t \cdot Z_t\big) >$ and

\begin{eqnarray}
\mathrm{d}\big(Y_t \cdot Z_t\big) & = & Y_t \cdot A Z_t + \widetilde{Y}_t \cdot A Z_t + \Big(\widetilde{Y}_t \cdot Z_t\Big) \mathrm{d}_{t} +
 \sigma_Y \mathrm{d}\widetilde{B}^2_{t},\\
                                               &=& \Big( Y_t  + \widetilde{Y}_t \Big)\cdot A Z_t  +  \Big(\widetilde{Y}_t \cdot Z_t\Big) \mathrm{d}_{t}+
 \sigma_Y \mathrm{d}\widetilde{B}^2_{t},\\
 &=& \kappa\Big( \widehat{y} + \frac{Y_t - \kappa Y_t }{\kappa}  - \lambda_Y\Big)\cdot  A Z_t +  \Big(\widetilde{Y}_t \cdot Z_t\Big) \mathrm{d}_{t}+ \sigma_Y \mathrm{d}\widetilde{B}^2_{t}.
 \end{eqnarray}
 In the following section, we give closed formulae of entropic risk measures on commodities. The dynamic of these commodities are given 
 the sub-section~\ref{sec:EnergyDerivatives}.
 
\section{Entropic  risk measures on commodities}
\label{Comparative-study}

We  use an entropic risk measure to express the risk associated with a contingent claim. 
A typical entropic risk measure depends on the risk aversion of the decision maker. This approach is interesting because it depends on individual preferences and opens a door to the theory of economic behaviour in order to quantify the risk. A drawback is  the risk aversion parameter is usually hard to measure. 
 The entropic  risk  is also called the exponential premium in the insurance literature, see~\cite{Goovaerts2004}.

\begin{definition}
Let $\mathcal{S}$ be  the linear space of  measurable functions on $\Omega$, including constant functions. An entropic risk measure $e_{\gamma}$ is 
defined for all $\psi\in\mathcal{S}$ by:
\begin{equation}
\label{entro}
 e_{\gamma}(\psi)\defegal -
 \gamma \ln \EE_{\PP}\Big[   \exp\big(  -\frac{1}{\gamma}\psi   \big)  \Big], \quad \gamma \in (0;\infty),
\end{equation}
\end{definition}
where $\PP$ is the historic probability distribution of $\psi$ and $\gamma$ a parameter of the entropic risk measure. Note that, this risk measure is convex but not coherent as it does not satisfy the positive homogeneity.
The entropic risk measure $e_{\gamma}$ corresponds to the certainty equivalent of $\psi$ when the utility function  considered is the exponential utility function with an Arrow-Pratt index $\gamma$;
see Gollier~\cite{Gollier2001} for the definitions  of certainty equivalent and Arrow-Pratt index. Also, $\gamma$ can be interpreted as the risk aversion parameter.  For a given contingent claim, we compute the associated risk when using the entropic risk  measure~\eqref{entro}. We shall focus on the three commodities claims
 described in Section~\ref{sec:EnergyDerivatives}. 

\subsection{Entropic risk measures on spot prices}
\begin{lemma}
\label{lemma1}
Suppose that the contingent claim $S_{t}$ belongs to $\mathcal{S}$ and  is given by 
\begin{equation}
\label{case1}
S_{t}=f(X_{t},Z_{t})\defegal X_{t} < \delta,Z_{t} >, \quad  t=1,\ldots,T,\quad\mbox{where}\quad \delta\defegal(\delta_{1},\ldots,\delta_{N})\in\RR^{N}.
\end{equation}
For $0\leq s< T $  the entropic risk at time $T$ knowing $Z_{s}$ is given by:
\begin{equation}
 e_{\gamma}(S_{T}|X_{s},Z_{s})= - <\lambda(s),Z_{s}>, \;\mbox{for} \;0\leq s<  T,
\end{equation}
where
$$ \lambda_{i}(s)\defegal  \gamma \ln<e^{A^{\star}(T-s)}\phi(s),e_{i}>, \quad  
\lambda(s)=(\lambda_{1}(s),\ldots,\lambda_{N}(s)), $$

$$ \phi_{i}(s)\defegal \exp\Big(- \frac{1}{\gamma}\delta_{i} \mu_{s,T}+
\frac{\delta_{i}^{2}\sigma_{s,T}^{2} }{2\gamma^{2}} \Big),  \quad  
\phi(s)=(\phi_{1}(s),\ldots,\phi_{N}(s)).  $$

%$$\mu_{t}=x_{0}e^{-\alpha t } + \beta(1-e^{-\alpha t})
  % \;\mbox{  and }\; 
%\sigma_{t}=  \frac{\sigma^{2}}{2\alpha}(1-e^{-2\alpha t}).
%$$
\end{lemma}

\begin{preuve}
Given $(X_{0},Z_{0})$, we wish to calculate the entropic risk associated with
 $S_{T}$.

\begin{eqnarray}
e_{\gamma}(S_{T}| X_{0},Z_{0})&=& - 
\gamma \ln \EE_{\PP}\Big[   \exp\big(  -\frac{1}{\gamma}S_{T} \big)  | X_{0},Z_{0}  \Big] \\
&=& -
\gamma \ln \EE_{\PP}\Big[   \exp\big(  -\frac{1}{\gamma}X_{T} < \delta ,Z_{T}>\big)  | X_{0},Z_{0}  \Big] \,
\mbox{ by }~\eqref{case1} \\
&=& -
\gamma \ln \EE_{\PP}\Big[  \EE_{\PP}\big[  \exp\big(  -\frac{1}{\gamma}X_{T} < \delta ,Z_{T} >\big)
|X_{0},Z_{0},Z_{T}\big]  | X_{0},Z_{0}  \Big] .
\label{risk1}
\end{eqnarray}
Define $\varepsilon_{i}\defegal \mathrm{exp}\big(  -\frac{1}{\gamma}X_{T}\delta_{i}  \big)$
and $\varepsilon\defegal\big( \varepsilon_{1},\ldots,\varepsilon_{N}\big)'$. Then~\eqref{risk1} becomes
 \begin{equation}
 e_{\gamma}(S_{T}| Z_{0})= - \gamma \mathrm{ln} \EE_{\PP}\Big[  \EE_{\PP}\big[ <\varepsilon,Z_{T}>
 |X_{0},Z_{0},Z_{T}\big]|X_{0},Z_{0}\Big].
 \end{equation} 
$X_{T}$ is  Gaussian with mean $\mu_{0,T}=x_{0}e^{-\alpha T } + \mu(1-e^{-\alpha T})$ and  variance 
$\sigma_{0,T}=  \frac{\sigma^{2}}{2\alpha}(1-e^{-2\alpha T})$. 
So
$$\EE_{\PP}\Big[  \varepsilon_{i}|Z_{0}\Big]=  \exp\Big(- \frac{1}{\gamma}\delta_{i} \mu_{0,T}+
\frac{\delta_{i}^{2}\sigma_{0,T}^{2} }{2\gamma^{2}} \Big)   \defegal\phi_{i}.
$$
Let us define $\phi\defegal\big( \phi_{1},\ldots,\phi_{N} \big)'$. From~\eqref{markovchaine}, $\EE\Big[ Z_{T}|Z_{0}\Big]=e^{AT}Z_{0}$.
So
\begin{eqnarray}
 e_{\gamma}(S_{T}| X_{0},Z_{0})&=&- \gamma \ln\EE_{\PP}\Big[ <\phi,Z_{T}>|Z_{0}\Big] = 
- \gamma \ln<\phi,e^{AT}Z_{0}>\nonumber,\\
 &=& - \gamma \ln<e^{A^{\star}T}\phi,Z_{0}>.
% \;\mbox{by}~\eqref{markovchaine}.
\end{eqnarray}
Write $\lambda_{i}(T)=\gamma \ln<e^{A^{\star}T}\phi,e_{i}>$, $\lambda\defegal(\lambda_{1},\ldots,\lambda_{N})'$,
\begin{equation}
 e_{\gamma}(S_{T}| X_{0},Z_{0})= - <\lambda(T),Z_{0}>.
\end{equation}

In the same way, for $0\leq s<  T $  the entropic risk at time $T$ knowing $Z_{s}$ is given by:

\begin{equation}
 e_{\gamma}(S_{T}| X_{s},Z_{s})= - <\lambda(s),Z_{s}>, \;\mbox{for} \;0\leq s< T,
\end{equation}
where
$$ \lambda_{i}(s)\defegal  \gamma \ln<e^{A^{\star}(T-s)}\phi(s),e_{i}>, \quad  
\lambda(s)=(\lambda_{1}(s),\ldots,\lambda_{N}(s))', $$

$$ \phi_{i}(s)\defegal \exp\Big(- \frac{1}{\gamma}\delta_{i} \mu_{s,T}+
\frac{\delta_{i}^{2}\sigma_{s,T}^{2} }{2\gamma^{2}} \Big),  \quad  
\phi(s)=(\phi_{1}(s),\ldots,\phi_{N}(s))',  $$

%$$\mu_{s,t}=  x_{0}e^{-\alpha (t-s) } + \beta(1-e^{-\alpha (t-s)})   \;\mbox{  and }\; 
%\sigma_{s,t}= \frac{\sigma^{2}}{2\alpha}(1-e^{-2\alpha (t-s)})  .$$

\end{preuve}

\subsection{Entropic risk measures on future contracts}

\begin{lemma}
\label{lemma2}
Suppose that the contingent claim $S_{t}$  is a forward contract given by
\begin{equation}
\label{fwd1}
F_{t}=f(X_{t},Z_{t})= e^{Y^{Z}_t(c_t,r_t)(t-T)}  X_{t}\defegal   e^{(r+y)(t-T)} X_{t} < \delta,Z_t>, \quad  t=1,\ldots,T.
\end{equation}
%where the constant $d,\,c\in\RR$ and $\delta\in\RR^{N}$.
For  $0\leq s<  T$, the  entropic risk at time $t$ knowing $Z_{s}$ is given by:

\begin{equation}
 e_{\gamma}(F_{T}| X_{s},Z_{s})=  - <\Lambda(s),Z_{s}>, 
\end{equation}
where
$$ \Lambda_{i}(s)\defegal  
\gamma \ln<e^{A^{\star}(T-s)}\Phi(s),e_{i}>
, \quad  
\Lambda(s)=(\Lambda_{1}(s),\ldots,\Lambda_{N}(s))', $$

%$$
%\psi_{i}(s)\defegal \exp\big(  -\frac{1}{\gamma}\delta_{i}d e^{c(s-t)}\big)
%$$

$$ \Phi_{i}(s)\defegal \exp\Big(- \frac{1}{\gamma}\delta_{i} e^{-(r+y)(T-s)}\mu_{s,T}+
\frac{\delta_{i}^{2}\sigma_{t}^{2} }{2\gamma^{2}}e^{-2(r+y)(T-s)} \Big),  \quad  
\Phi(s)=(\Phi_{1}(s),\ldots,\Phi_{N}(s))',  $$

%$$\overline{\mu}_{t}=e^{(\alpha-c) (t-s)}\Big(  x_{0} -\alpha\beta  \int_{s}^{t}e^{-\alpha u} \mathrm{d}u \Big)     \;\mbox{  and }\; 
%\overline{\sigma}_{t}=  e^{2 (\alpha-c) (t-s)} \sigma^{2}\int_{s}^{t}e^{-2 \alpha u} \mathrm{d}u .$$

\end{lemma}

\begin{preuve}

Given $(X_{0},Z_{0})$, we wish to calculate the entropic risk associated with
 $F_{T}$ defined in~\eqref{fwd1}.
\begin{eqnarray}
e_{\gamma}(F_{T}| X_{0},Z_{0})&=& -
\gamma \ln \EE_{\PP}\Big[   \exp\big(  -\frac{1}{\gamma}F_{T} \big)  | X_{0},Z_{0}  \Big] \\
&=& -
\gamma \ln \EE_{\PP}\Big[   \exp\big(  -\frac{1}{\gamma}X_{T} e^{- (r+y)T} < \delta ,Z_{T}>\big)  | X_{0},Z_{0}  \Big] \,
\mbox{ by }~\eqref{case1}\nonumber \\
&=& -
\gamma \ln \EE_{\PP}\Big[  \EE_{\PP}\big[  \exp\big(  -\frac{1}{\gamma}X_{T} e^{-(r+y)T}  < \delta ,Z_{T} >\big)
|X_{0},Z_{0},Z_{T}\big]  | X_{0},Z_{0}  \Big]  \label{risk2}.
%&= &-\underbrace{\gamma \ln \EE_{\PP}\Big[  \EE_{\PP}\big[  \exp\big(  -\frac{1}{\gamma} d  e^{-cT}  < \delta ,Z_{T} >\big)|X_{0},Z_{0},Z_{T}\big]  | X_{0},Z_{0}  \Big] }_{ (\star)}  \\
%&&-\underbrace{\gamma \ln \EE_{\PP}\Big[  \EE_{\PP}\big[  \exp\big(  -\frac{1}{\gamma}e^{-cT}X_{T}   < \delta ,Z_{T} >\big)|X_{0},Z_{0},Z_{T}\big]  | X_{0},Z_{0}  \Big] }_{ (\star\star)}\label{risk2}
\end{eqnarray}
%From ($\star$), we have
%\begin{eqnarray}
  %\EE_{\PP}\big[  \exp\big(  -\frac{1}{\gamma} d  e^{-cT}  < \delta ,Z_{T} >\big)|X_{0},Z_{0},Z_{T}\big] &=&
      %    \EE_{\PP}\Big[  \exp\big( <  - \frac{1}{\gamma} \delta  d  e^{-cT} ,Z_{T}> \big)| Z_{0}\Big] \\
          %&=&<\exp\big(  - \frac{1}{\gamma} \delta  d  e^{-cT}\big) ,e^{AT}Z_{0}>
%\end{eqnarray}
%So, $(\star)=\gamma\ln <  \exp\big(  - \frac{1}{\gamma} \delta  d  e^{-cT}+ A^{\star}T\big)     ,Z_{0}>$.\\ 

%Now, let us focus on $(\star\star)$. 
We will follow the same steps as in the proof of Lemma~\ref{lemma1}.
Define $\varepsilon_{i}\defegal \mathrm{exp}\big(  -\frac{1}{\gamma}X_{T}e^{-(r+y)T}\delta_{i}  \big)$
and $\varepsilon\defegal\big( \varepsilon_{1},\ldots,\varepsilon_{N}\big)'$. Then~\eqref{risk2} becomes
\begin{equation}
e_{\gamma}(F_{T}| X_{0},Z_{0})= -  \gamma \ln \EE_{\PP}\Big[  \EE_{\PP}\big[<\varepsilon, Z_T>| X_0,Z_0,Z_T\big]|X_0,Z_0\Big]
\end{equation}
Now, evaluate $ \EE_{\PP}\big[<\varepsilon, Z_T>| X_0,Z_0,Z_T\big]$. We know that $X_T$ is a Gaussian with parameters $\mu_{0,T}$ and $\sigma_{0,T}$. Then
\begin{equation}
 \EE_{\PP}\big[<\varepsilon_i| Z_0|\big]=
 \exp\Big[ -\frac{1}{\gamma}e^{-(r+y)T}\delta_i \mu_{0,T} + \frac{\delta_i^2\sigma^2_{0,T}}{2\gamma^2}e^{-2(r+y)T}   \Big]\defegal \Phi_i(0).
\end{equation}
It follows that
%\begin{equation*}
$e_{\gamma}(F_{T}| X_{0},Z_{0})= - <\Lambda(T),Z_{T}>$, where
%\end{equation*}
 $$\Lambda_{i}(T)\defegal   \gamma\ln <e^{A^{\star}T} \Phi_{i}(T),e_{i} > ,\quad  
 \Lambda(T)=(\Lambda_{1}(T),\ldots,\Lambda_{N}(T))', 
$$
 $$\Phi_{i}(T) \defegal \exp\Big(- \frac{1}{\gamma}\delta_{i} e^{-(r+y)T}\mu_{0,T}+
\frac{\delta_{i}^{2}\sigma_{0,T}^{2} }{2\gamma^{2}}e^{-2(r+y)T} \Big), \quad \Phi(T)=(\Phi_{1}(T),\ldots,\Phi_{N}(T))'\,.$$

%$$ \overline{\mu}_{T}=  e^{(\alpha-c) T}\Big(  x_{0} -\alpha\beta  \int_{0}^{T}e^{-\alpha u} \mathrm{d}u \Big)     \;\mbox{and}\;    \overline{\sigma}_{T}   =e^{2 (\alpha-c) T} \sigma^{2}\int_{0}^{T}e^{-2 \alpha u} \mathrm{d}u.    $$

Following the same calculation  as in the proof of Lemma~\ref{lemma1}, we obtain the desired result at time $s<T$.
\end{preuve}

\subsection{Entropic risk measures on commodity swaps}

%\begin{lemma}
Assume that the contingent claim is a commodity swap given by
\begin{equation}
\label{eq:ercwp}
W_T= f(X_t, Z_t)=\displaystyle \sum_{t=1}^{T} e^{-r_t}X_t\Big( e^{Y^{Z}_t(c_t,r_t)(t-T)}   - 1 \Big).
\end{equation}
%\end{lemma}
For $0\leq s< T $  the entropic risk at time $T$ knowing $Z_{s}$ is given by:
\begin{equation}
\label{eq:fswpeq}
 e_{\gamma}(W_T|X_{s},Z_{s})= - \gamma \ln \EE_{\PP} \Big[ \psi(s) 
\exp\Big(-\frac{1}{\gamma}e^{- r_s}X_s \Big( e^{Y_t^Z(c_s,r_s)(s-T)} -1 \Big) \Big)| X_s,Z_s \Big], \;\mbox{for} \;0\leq s<  T,
\end{equation}
where $\psi(s) =\displaystyle
\prod_{t=1}^{s-1}  \EE_{\PP} \Big[  \exp\Big(-\frac{1}{\gamma}e^{- r_s}X_s \big( e^{Y_t^Z(c_s,r_s)(s-T)} -1\big)
 \Big)|X_{s-1}, Z_{s-1} \Big]$. 
The formula obtained in \eqref{eq:fswpeq} is not complete as the ones obtained in Lemmas \ref{lemma1} and \ref{lemma2}. 
Thus finding the entropic risk associated with the cash flows  exchanged at the maturity will require computing conditional expectations with respect to all the stochastic processes involved in the formula, at each time period. To stay in the context of this study and for the sake of simplicity, we will not go further on finding simpler closed formula of the entropic risk measure associated with the commodities swaps.
Next, we shall illustrate and discuss only the entropic risk measures associated with linear spot prices and future contracts.

\section{Illustration}
\label{Sec:Illust}
Consider the historical prices of the WTI crude oil for a period of one year. We consider the opening price from $ 09/30/2019$ to
 $08/31/2020$\footnote{Data are obtained from https://markets.businessinsider.com/commodities/oil-price}. The data points are 
used as they are. Thus, missing data from none opening days or omitted are not approximated. Below is a chart of the historical prices.
\begin{figure}[htbp]
\begin{center}
\includegraphics[width=12cm,height=8cm]{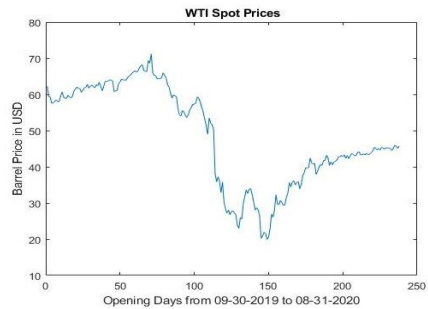} %{HistoricalSpotPrices.pdf}
\end{center}
\caption{Historical data of Crude Oil Spot Prices - Opening Days from  09/30/2019 to 08/31/2020.\label{chart1}}
\end{figure}
If we look closely the graph, we can observe that the dynamic of the spot price is in three phases during the period of this observation.
The first three  months are characterized  by a slight upward trend, followed by a sharp downturn for another semester and a slight upward trend again. 

\subsection{Parameters determinants and simulations}
We are observing a two state economy: sharp downturn and slight recovery, amid of the global recession around the word. Since the economic regime  lasts over a certain period of time, we can use a four stage discrete Markov chain to model the state of the economy. 
Choose a transition probability matrix $A$ as follows:
\begin{equation*}
A = \left(
\begin{array}{cccc}
0.75 &  0.25 & 0 & 0 \\
0.25 & 0.75 & 0 & 0 \\
0      &  0     & 0.25 & 0.7\\
0     &  0      & 0.75 & 0.25
\end{array}
\right) 
\end{equation*}
\begin{figure}[!hbp]
\begin{center}
\includegraphics[width=7cm,height=5cm]{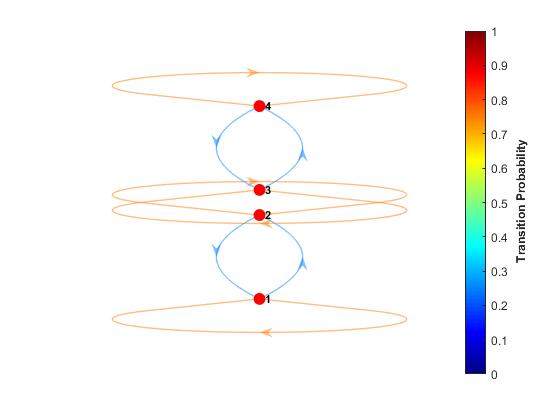}% {TransitionMatrix.jpg}%{FourStageTransitionMatrix.pdf}
\end{center}
\caption{Four Stage Transition Matrix\label{A_Matrix}}
\end{figure}
%\newpage
That is, if the economy is in downturn at time $t$, the probability that it stays in recession is $0.75$.
If the economy is growing, the probability that it keeps growing is $0.25$. The zero entries correspond to a stagnation situation, where the economy is staying steady without any significant movement. The Figure~\ref{A_Matrix} is an illustration of these probabilities.

Based on the dynamic of the observed prices, we can calibrate the OU spot prices. %That is not our main concern in this study.
Below is a simulated spot price.
\begin{figure}[htbp]
\begin{center}
\includegraphics[width=8cm,height=9cm]{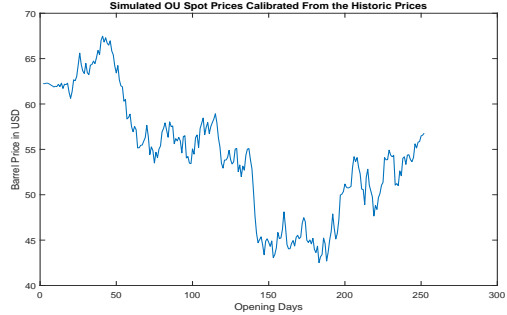}
\end{center}
\caption{Simulated WTI Crude Oil Spot Prices\label{Sim}}
\end{figure}

Now, we can simulate the dynamic entropic risk associated with each energy derivative product.  The parameters of the dynamatic of the spot price are: 
\begin{table}[!htbp]
\begin{center}
\begin{tabular}{|c|c|c|c|}
\hline
\cline{1-4}
\multicolumn{4}{c}{Parameters of the Dynamic of the Spot Price}\\
\hline
$\alpha$    & $\mu$         & $\gamma$ &Initial Spot Price $S_0$ \\
\hline
$5$            & $\$ 48.22$ & $\$ 13.66$ &  $ \$ 62.24$\\
\hline\hline
\end{tabular}
\caption{Parameters of the OU Process.}
\end{center}
\end{table}

\subsection{Sensitivity analysis based on the risk aversion parameter and the time horizon}
\begin{figure}[!htbp]
\begin{center}
\includegraphics[width=8cm,height=7cm] {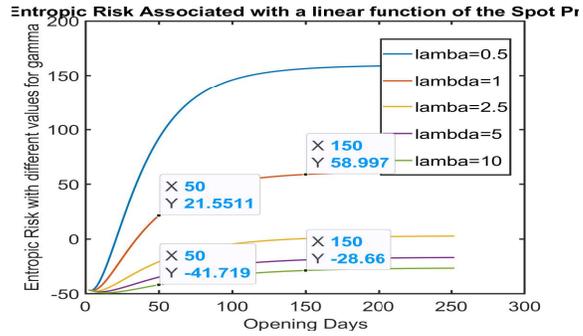}%{FourStageTransitionMatrix.pdf}LinearEntropic1.pdf
\end{center}
\caption{Entropic risk of a  linear derivative for different values of gamma\label{Linear}}
\end{figure}

For different value of $\gamma$, Table~\ref{Table1} presents the value of the entropic risk. Note that the entropic risk does not have an interpretation in term of capital at risk. Figure~\ref{Linear} shows the computed entropic risk for different values of the risk measure parameter $\gamma$. 
We assume that $\delta_i=0.75$, which implies a price decrease of $25\%$. Next, we evaluate the entropic risk associated with the linear derivative for different level of risk aversion.
 Figure~\ref{Linear} shows the values of the simulated entropic risk. By definition, these vales decrease with respect to $\gamma$. However, these results show that the magnitude of the difference is smaller  when $\gamma$ increases. Next, we shall do a quick analysis for some specific values in the Table~\ref{Table1}.

\begin{table}[!htbp]
\begin{center}
\begin{tabular}{|p{3cm}|*{4}{c|}}
\cline{2-5}
\multicolumn{1}{c|}{}
 & \multicolumn{4}{c|}{\textbf{Risk  Aversion}} \\[3mm]
\cline{2-5}
\multicolumn{1}{c|}{\textbf{Horizon}} & $\gamma=1$ & $\gamma=2.5$  &  $\gamma=5$   &  $\gamma=10$  \\[3mm]
\hline
\bfseries $T=50$ days &  21.55 &  -20.68 & -34.74  & - 41.72  \\[3mm]
\hline
\bfseries $T=150$ days & 60 &  0.52 & -18.97   & - 28.66  \\[3mm]
\hline
\bfseries Variation ($\%$) & 178.42  & 102.51   & 45.39   & 31.30   \\[3mm]
\hline\hline
\end{tabular}
\caption{Sensitivity  analysis with respect to the risk aversion parameter and the time horizon.\label{Table1}}
\end{center}
\end{table}

\begin{figure}[!htbp]
\begin{center}
\includegraphics[width=8cm,height=7cm] {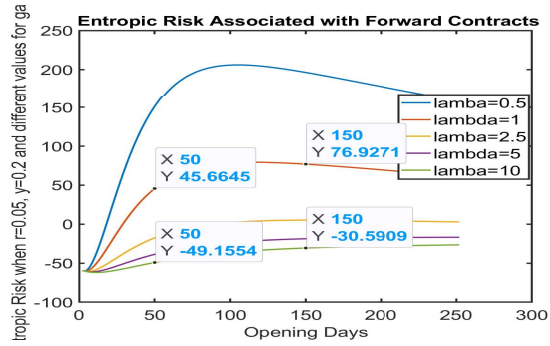}%{FourStageTransitionMatrix.pdf}LinearEntropic1.pdf
\end{center}
\caption{Entropic Risk of Forward Prices for different values of gamma\label{Forward}}
\end{figure}

The value of the entropic risk decreases with respect to the risk aversion. This is related to the expression of the entropic risk measure itself. 
The determinants and estimations of the risk aversion parameter are studied in~\cite{RiskAverionDet}. Usually, the value of the risk aversion depends on the decision maker's behavior. In Seck et~\emph{al.},~(\cite{Seck2011}), it is shown that under the assumption of some particular formulation of a class of risk measures, maximizing a profit subject to these risk constraints is equivalent to maximizing a certain class of multi-attribute utility functions. These multi-attribute utility functions express loss aversion instead of risk aversion. Similar results are obtained recently~\cite{RiskAverionDet}. Thus, the value of the risk aversion parameter depends on the distribution of the underlying asset, which is somehow another way to express the decision maker risk preferences.  Another factor that is included in the Table~\ref{Table1} is the time horizon. We can see that, for any values of the risk aversion, the entropic risk associated with the underlying asset decreases as time elapses.  To conclude, we are seeing a form of time consistency of the preferences regardless the state of the economy. We observe the same behavior of the entropic risk measure if the underlying assets is a forward contract, see Figure~\ref{Forward} and Table~\ref{Table3}.

\begin{table}[!htbp]
\begin{center}
\begin{tabular}{|p{3cm}|*{4}{c|}}
\cline{2-5}
\multicolumn{1}{c|}{} 
 & \multicolumn{4}{c|}{\textbf{Risk  Aversion}} \\[3mm]
\cline{2-5}
\multicolumn{1}{c|}{\textbf{Horizon}} & $\gamma=1$ & $\gamma=2.5$  &  $\gamma=5$   &  $\gamma=10$  \\[3mm]
\hline
\bfseries $T=50$ days & 45.66   & -17.60  & -38.67  & -49.16  \\[3mm]
\hline
\bfseries $T=150$ days & 76.93 & 5.19  & -18.69  & -30.59  \\[3mm]
\hline
\bfseries Variation ($\%$) & 68.48  & 129.49  & 51.67   & 37.77   \\[3mm]
\hline\hline
\end{tabular}
\caption{Sensitivity analysis with respect to the risk aversion parameter and the time horizon.\label{Table3}}
\end{center}
\end{table}
%Curiously, if the underlying asset are forward contracts, the  values of the entropic risk are not decreasing

\subsection{Sensitivity analysis based on the convenience yield}
\begin{figure}[!htbp]
\begin{center}
\includegraphics[width=8cm,height=7cm] {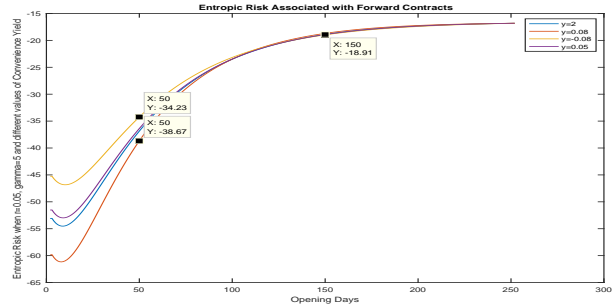}%{FourStageTransitionMatrix.pdf}LinearEntropic1.pdf
\end{center}
\caption{Entropic risk of Future Crude Oil with respect to the convenience yield over time\label{FutureConvYield}}
\end{figure}

Some historical data analysis has shown that oil futures have reflected a convenience yield of 8\% per year on average, from January 1986 to May 2008\footnote{Refer to the analysis  in https://www.interfluidity.com/posts/1214354098.shtml}.  The convenience yield is zero or negative if future markets are discouraging storage. This happens recently at the beginning of the Covid-19 health crisis for crude oil commodities; this phenomenon is known as  ''backwardization''. Recent study has shown statistical evidence of jump bevahior of the convenience yield in relation to the market conditions such as investment on storage capacity, leasing facility or drilling, see \cite{ConvenienceYield2020}. Next, we analyze the sensitivity of the entropic risk with respect to the convenient yield and the time horizon. Figure~\ref{FutureConvYield} provides the rends of the entropic risk for different value of the convenience yield when time elapses. It turns out that for different value of the convenience yield, it will arrive a moment where the entropic risk converge to an equilibrium value. This can be interpreted as a mean-reverting and a term-structure behavior of the convenience yield. That is the behavior of the convenience yield over a short period time might differ over a long period of time. Now, in this context a short period of time could be the order of two to three months. Further studies might determine whether these behaviors of the convenience yield are actually observable in the crude oil market.  

\section{Conclusion}
This paper introduces a new class of risk measures that can take into account the decision maker's preference and the state of the economy.
The tractability of these risk measures are studied through simulations on crude oil futures. The results are promising as they show the term structure
and the mean-reverting behavior of the convenience yield.  However, the determination of the risk aversion parameter in the setting of this paper remains an open topic and requires further studies.
%\nocite{Schwartz2000}\nocite{Pauline2004}\nocite{Mania2005}\nocite{Detlefsen2005}\nocite{Delbaen2006}
%\nocite{Follmer2006}\nocite{Delbaen2006three}

%\bibliographystyle{gENO}
%\begin{thebibliography}{}

%\newpage
%\bibliographystyle{apa}
\bibstyle{apa}

%\bibstyle{apa}

%\bibliographystyle{amsplain}
\bibliography{Seck_Elliott2_Dec10_2020}

%\end{thebibliography}

\end{document}